\documentclass[letter]{aa} 
\usepackage{graphicx}
\begin{document}
   \title{Time variation of radial gradients in the galactic disk: electron 
    temperatures and abundances}

   \author{W. J. Maciel
          \inst{1},
          C. Quireza
          \inst{2},
          \and
          R. D. D. Costa\inst{1}
          }

   \offprints{W. J. Maciel}

   \institute{Instituto de Astronomia, Geof\'\i sica e Ci\^encias Atmosf\'ericas
    (IAG/USP), Universidade de S\~ao Paulo, Rua do Mat\~ao 1226, 05508-900 S\~ao
    Paulo SP, Brazil\\
              \email{maciel@astro.iag.usp.br, roberto@astro.iag.usp.br}
         \and
             Observat\'orio Nacional, Rua General Jos\'e Cristino 77, 20921-400
             Rio de Janeiro RJ, Brazil\\
             \email{quireza@on.br}
                   }

   \date{Received ..., ...; accepted ..., ...}

 
  \abstract
   {}
   {We investigate the electron temperature gradient in the galactic 
   disk as measured by young HII regions on the basis of radio 
   recombination lines and the corresponding gradient in planetary 
   nebulae (PN) based on [OIII] electron temperatures. The main goal 
   is to investigate the time evolution of the electron temperature 
   gradient and of the radial abundance gradient, which is essentially 
   a mirror image of the temperature gradient.}
   {The recently derived electron temperature gradient from radio 
   recombination lines in HII regions is compared with  a new determination 
   of the corresponding gradient from planetary nebulae for which the 
   progenitor star ages have been determined.}
   {The newly derived electron temperature gradient for PN with progenitor 
   stars with ages in the 4-5 Gyr range is much steeper than the corresponding 
   gradient for HII regions. These electron temperature gradients are converted 
   into O/H gradients in order to make comparisons with previous estimates
   of the flattening rate of the abundance gradient.}
   {It is concluded that the O/H gradient has flattened out in the past
   5 Gyr at a rate of about 0.0094 dex kpc$^{-1}$ Gyr$^{-1}$, in good 
   agreement with our previous estimates.}

   \keywords{abundances -- gradients -- chemical evolution
               }
   \authorrunning{W. J. Maciel et al.}
   \titlerunning{Time variation of radial gradients}

   \maketitle
%

\section{Introduction}

Radial abundance gradients in the galactic disk and their time
variations are among the main constraints of chemical evolution
models for the Milky Way. These gradients can be determined from
a variety of objects, such as HII regions, cepheid variables,
open clusters and planetary nebulae (PN). In a recent series of 
papers, Maciel et al. (\cite{mcu}, \cite{mlc1}, \cite{mlc2})
estimated the time variation of the radial abundance gradients
taking into account a large sample of PN for which abundances
of O/H, S/H, Ne/H and Ar/H have been derived. Based on individual
estimates of the progenitor star ages, it was concluded that
the radial gradients are flattening out at an average rate of 
about $0.005-0.010\, {\rm dex}\, {\rm kpc}^{-1}\,{\rm Gyr}^{-1}$
for the last 8 Gyr, approximately. A comparison of the PN 
gradients with results from HII regions, OB stars and associations,
cepheids and, especially, open cluster stars, strongly
supports these conclusions.

On the other hand, it has long been known  that a positive 
electron temperature gradient of about $250-450\,$K/kpc is
observed in the galactic disk, mainly on the basis of radio 
recombination line work on HII regions (see for example 
Churchwell \& Walmsley \cite{cw}, Churchwell et al.
\cite{csmmh}, Shaver et al. \cite{shaver}, Wink et al.
\cite{wink}, Afflerbach et al. \cite{afflerbach}, and
Deharveng et al. \cite{deharveng}). Such a gradient 
is interpreted as a reflection of the radial abundance 
gradient of elements such as O/H, S/H, etc. in the galactic 
disk, since these elements are effective coolants of 
the ionized gas (see for example Shaver et al. \cite{shaver}).

Recently, Quireza et al. (\cite{cintia}) presented a detailed
study of a large sample containing over a hundred HII regions 
spanning about 17 kpc in galactocentric distances for
which accurate electron temperatures were determined from
radio recombination lines, specifically H91$\alpha$ and
He91$\alpha$. The observations were made with the 140 Foot
telescope of the National Radio Astronomy Observatory
(NRAO), and are of unprecedented sensitivity compared
with previous studies. According to this work, the best estimate
of the gradient, obtained from a sample of 76 sources
with high quality data, is $dT_e/dR \simeq 287 \pm 46\,$K/kpc, 
with no significant variations along the galactocentric
distances. A slightly larger gradient (up to 17\%) was
obtained by excluding some HII regions which are closer
to the galactic centre, and may not belong to the
disk population.

Regarding planetary nebulae, our earlier work (Maciel \&
Fa\'undez-Abans \cite{mfa}) based on a sample of PN 
classified according to the Peimbert types (cf. Peimbert \cite{mp}) 
suggested a positive electron temperature gradient in the 
range $550-800\,$K/kpc, somewhat steeper than the HII region 
gradients observed at the time. 

In this work, we take into account the recent PN samples 
analyzed by Maciel et al. (\cite{mcu}, \cite{mlc1}, \cite{mlc2}) 
and derive the PN electron temperature gradient for a sample
of objects having similar ages. A comparison of the obtained $T_e$ 
gradient with the recently derived value by Quireza et al. 
(\cite{cintia}) for HII regions gives then an independent 
estimate of the time variation of the radial abundance gradients 
in the galactic disk.


\section{The electron temperature gradient in the galactic disk}

The determination of abundance gradients is a difficult task,
basically for three main reasons. First, the magnitudes of the
gradients are small, amounting at most to a few hundredths
in units of dex/kpc, so that a relatively large galactocentric
baseline is needed in order to obtain meaningful results. Second,
the uncertainties both in the abundances and in the distances
contribute to the observed scattering, so that large samples
are usually needed. Third, chemical evolution models generally 
predict some time variation of the gradients, so that it is 
extremely important to take into  account in a given sample only 
objects with similar ages. For these reasons, some of the analyses
of gradients in the literature produce relatively flat gradients
(see for example Perinotto \& Morbidelli \cite{pm}). On the
other hand, accurate and homogeneous abundances eliminate some
of these problems, so that relatively steeper gradients are obtained,
as in Pottasch  \& Bernard-Salas (\cite{pbs}).

\begin{figure}
   \centering
   \includegraphics[angle=-90,width=8cm]{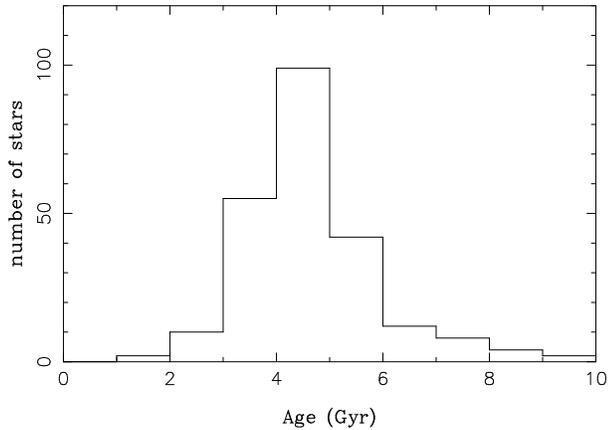}
      \caption{Age distribution of the PN progenitor stars
       in the Basic Sample of Maciel et al. (\cite{mlc2}).
              }
         \label{histog}
   \end{figure}

In our recent work, we made an attempt to overcome some of
these problems, and estimated the  individual ages of the PN 
progenitor stars using an age-metallicity
relationship which also depends on the galactocentric distance.
As a result, we have obtained the age distribution shown
in  Fig.~\ref{histog}, adopting our Basic Sample, which is the 
largest and most complete sample we have considered, containing 
234 nebulae (see Maciel et al. \cite{mcu}, \cite{mlc1}, \cite{mlc2}
for details).

%

\begin{figure}
   \centering
   \includegraphics[angle=-90,width=8cm]{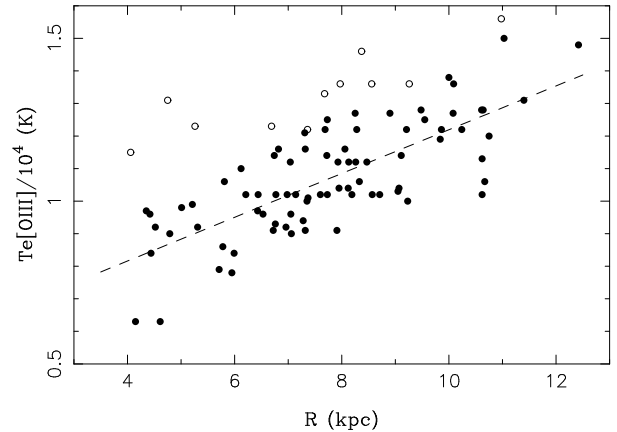}
      \caption{Galactocentric variation of the [OIII] 
      electron temperatures for PN with progenitor ages
      of 4--5 Gyr. The empty circles show some
       nebulae having extremely hot central stars,
       not included in the linear regression analysis.
              }
         \label{gradte}
   \end{figure}

It can be seen that the ages are strongly peaked around
4--5 Gyr, where we have 99 objects. In the present work, 
we will consider the objects in this age bracket, in order 
to make comparisons with the younger HII regions. We have 
then collected the electron temperatures of the planetary
nebulae, selecting only the [OIII] temperatures in order 
to keep our sample as homogenous as possible. These 
temperatures are determined from the ratio of the
[OIII] 4363/5007\AA\ lines, which are usually among
the brightest collisionally excited emission lines
in the spectra of planetary nebulae. We have 
preferred our own data where available (Costa et al.
\cite{costa1}, \cite{costa2}, see a list of 
references in Maciel et al. \cite{mcu}, \cite{mlc1},
\cite{mlc2}), with additional data by Henry et al. 
(\cite{henry}), Kingsburgh \& Barlow (\cite{kingsburgh}),
and Cahn et al. (\cite{cks}). The resulting $T_e$
variation with galactocentric distance $R$ is shown
in Fig.~\ref{gradte}, where we adopted the same distances
and solar galactocentric radius as in our previous work.
The total number of objects in Fig.~\ref{gradte} is somewhat
lower than shown in the 4--5 Gyr bracket of Fig.~\ref{histog},
as for a few nebulae we could not obtain accurate electron 
temperatures. 

It can be seen from Fig.~\ref{gradte} that there is
a clear tendency in the sense that higher electron
temperatures are associated with larger galactocentric
distances. The best derived $T_e$ gradient for this sample
of PN is $dT_e/dR \simeq 670\pm  65\,$K/kpc, with a correlation
coefficient of $r \simeq 0.76$, which is similar 
to the gradient for the `selected sample' of 
our earlier paper (Maciel \& Fa\'undez-Abans \cite{mfa}).
Adopting instead a homogeneous set of [OIII] electron
temperatures from Henry et al. (\cite{henry}), which
is the largest homogeneous sample available for
these nebulae, we obtain essentially the same 
result, namely $dT_e/dR = 680\pm  140\,$K/kpc and 
$r \simeq 0.65$, so that the correlation is real.
Our best derived slope is illustrated by the
dashed line in Fig.~\ref{gradte}.

The average uncertainty in the determination of the
electron temperatures is generally considered to
be within 10\% for the brightest nebulae, which 
corresponds roughly to 1000 K for most objects 
(see for example Kingsburgh \& Barlow \cite{kingsburgh}
and Krabbe \& Copetti \cite{krabbe}).
For the galactocentric distances an average uncertainty
is more difficult to establish, as it depends on
the adopted distances. Since most objects in
Fig.~\ref{gradte} are located within about 3 kpc
from the solar galactocentric radius, an average
error of 50\% in the distances would correspond
to a shift in the galactocentric distances of 
about 0.03 -- 1.0 kpc depending on the distance
and the direction of the line of sight to the
nebula. As a comparison, for the HII regions
in the sample by Quireza et al. ({\cite{cintia}),
average formal uncertainties in the electron temperatures
are within 2\%, but systematic errors may increase this 
uncertainty up to 10\% for  the best data. For 
spectrophotometric distances, average errors of 
15\% are quoted, while for kinematic distances non-circular
streaming motions may increase this figure to about 25\%. 
From the $T_e \times R$ plot by
Quireza et al. ({\cite{cintia}), an average
dispersion of about 2200~K can be obtained,
which is about half the dispersion in 
Fig.~\ref{gradte}.

It should be noted that the dispersion observed in 
Fig.~\ref{gradte} is probably real, since the electron
temperatures may be affected by several factors, such
as differences in the effective temperature of the central 
stars, presence of dust, optical depth effects, 
electron density and temperature fluctuations, etc., apart from 
the main cause of the $T_e \times R$ variation, namely, 
the radial abundance gradient. These effects are also
partially responsible for the  observed dispersion
in HII regions, but for PN the variations in the
effective temperatures of the central stars and the
uncertainties in the distances are larger, so that
the observed dispersion in Fig.~\ref{gradte} is larger
than in the case of HII regions. As a consequence, some 
nebulae appear not to follow the observed correlation
very closely. In particular, PN having extremely
hot central stars, with temperatures in excess of
$10^5\,$K, generally have higher electron temperatures
than expected by their galactocentric distances.
Furthermore, these objects come from more massive
progenitors than most nebulae, so that their ages may be 
lower than 4--5 Gyr, as assumed. Some examples include
M1-57, Me2-1, PB6, NGC 6620 and a few others, which
are plotted in the figure as empty circles. Other objects
with hot central stars, such as NGC 2899, NGC 6302, 
NGC 6537 and NGC 7008 are not plotted in Fig.~\ref{gradte}, 
as their electron temperatures are too high to fit the
scale. All these nebulae have not been included in the 
linear regression, so that our derived electron temperature
gradient applies to stars with temperatures lower than
$10^5\,$K. In this analysis we have used Zanstra 
temperatures and energy-balance temperatures (see for
example Preite-Martinez et al. \cite{andrea}, M\'endez et al. 
\cite{mendez}, Zhang \cite{zhang}, and Stasi\'nska et al. 
\cite{stasinska}).

Another interesting object is M1-9, 
which is the nebula with the largest
galactocentric {\bf distance} in the sample ($R \simeq 12.4\,$kpc).
In view of its position on the $T_e \times R$ plane,
it may single-handedly affect the derived slope. For this object, our
own results suggest an electron temperature of
$T_e \simeq 11000\,$K (Costa et al. \cite{costa2}),
but a more detailed study by Tamura \& Shibata 
(\cite{ts90}) and Shibata \& Tamura (\cite{st85})
gives a larger value, $T_e \simeq 14800\,$K,
which is adopted here. This object may then alter
the derived slope by about 50 K/kpc, but the main
conclusions of this paper are unaffected.

The association of higher electron temperatures with
lower metallicities can be seen from Fig.~\ref{teoh},
where we show the inverse correlation between the
[OIII] electron temperatures and the O/H abundances 
for the objects with ages in the 4--5 Gyr bracket. 
PN with central stars hotter than $10^5\,$K
are also shown as empty circles.
Again a relatively large dispersion is observed, 
but the inverse correlation is clear, confirming 
that oxygen is among the main coolants in the
photoionized gas within the planetary nebulae.

 \begin{figure}
   \centering
   \includegraphics[angle=-90,width=8cm]{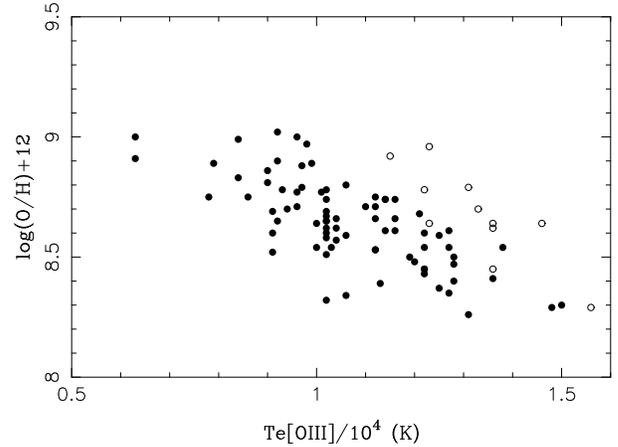}
      \caption{The inverse correlation between the 
       [OIII] electron temperatures and the O/H
       abundances for PN in the 4--5 Gyr 
       age bracket. The empty circles show some
       nebulae having extremely hot central stars.
              }
         \label{teoh}
   \end{figure}

%

\section{Comparison of the PN and HII gradients}

From a straigthforward comparison of the electron temperature gradient
for PN and HII regions we can already conclude that there is some 
flattening of the gradients during the time interval
of about 5 Gyr, which is essentially the difference between the
ages of the PN progenitor stars considered in this work and 
the much younger HII regions in the sample by Quireza et al.
(\cite{cintia}). In other words, the conclusions of our recent
series of papers are supported by the obtained differences between
the electron temperature gradients of PN and HII regions,
in view of the fact that these gradients essentially reflect
the radial abundance gradients in the galactic disk.
In order to make a direct comparison
with the estimated flattening rate of the abundance gradients
derived by Maciel et al. (\cite{mlc1}), we can convert the
$T_e$ gradient into the equivalent O/H gradient. For HII regions
we can use the calibration by Shaver et al. (\cite{shaver}),
according to which the oxygen gradient is related to the
electron temperature gradient by

   \begin{equation}
   {d \log({\rm O/H}) \over dR} \simeq -1.49 \times 10^{-4} \ 
   {d T_e \over dR} \,
   \end{equation}

\noindent
where the temperature gradient is in K/kpc and the oxygen gradient
is in dex/kpc. This relation is also supported by more recent work
on HII regions, such as the analysis by Deharveng et al. (\cite{deharveng}). 
The oxygen gradient for HII regions is then 
$d\log({\rm O/H})/dR \simeq -0.043\,$dex/kpc,
which is similar to the value obtained by Deharveng et al. (\cite{deharveng})
based on an entirely different sample. Also, an O/H gradient
of $-0.044$ dex/kpc was recently obtained by Esteban et al.
(\cite{esteban}) from oxygen recombination lines, a method almost
independent of the assumed electron temperatures, and totally
independent of the [OIII] forbidden lines. A similar estimate has been
presented by Quireza et al. (\cite{cintia}), also based on the Shaver
et al. ({\cite{shaver}) calibration. 

For planetary nebulae, we can have an idea of the corresponding O/H 
gradient by inspecting Table 1 of Maciel et al. ({\cite{mlc1}) for 
Group II objects (ages of 4--5 Gyr), from which we get 
$d\log({\rm O/H})/dR \simeq -0.089 \pm 0.003\,$dex/kpc. 
Alternatively, we can compute the O/H gradient directly for the PN sample 
adopted here, in which case we get a similar value, $d\log({\rm O/H})/dR 
\simeq -0.090\,$dex/kpc. Therefore, the flattening rate of the oxygen 
gradient can be estimated by

   \begin{equation}
   \chi \simeq {1 \over \Delta T} 
   \left[{d \log({\rm O/H}) \over dR}\Big\arrowvert_{HII} 
   - {d \log({\rm O/H}) \over dR}\Big\arrowvert_{PN}\right]
    \,
   \end{equation}

\noindent
so that $\chi \simeq 0.0094\, {\rm dex}\, {\rm kpc}^{-1}\, {\rm Gyr}^{-1}$
where we have used $\Delta t \simeq 5\,$Gyr. In view of our discussion
on the electron temperatures of planetary nebulae, we would expect
the rate to be somewhat smaller than this value, which is
in excellent agreement with the range estimated by Maciel et al. (\cite{mlc1}),
namely $\chi \simeq 0.005-0.010\  {\rm dex}\, {\rm kpc}^{-1}\, {\rm Gyr}^{-1}$.

As mentioned in the introduction, abundance gradients and
their time variation are valuable constraints for chemical
evolution models. As an illustration, we have compared our
derived flattening rate with the predictions of some recently
published models for the Milky Way. As discussed by Maciel
et al. (\cite{mlc2}) there may be large discrepancies between
different chemical evolution models, even whithin the 
so-called `inside-out' class of models. In particular,
Hou et al. (\cite{hou}) adopted an exponentially decreasing 
infall rate for the galactic disk, in which a rapid increase in 
the metal abundance at early times in the inner disk leads 
to a steep gradient. As times goes on, the star formation migrates 
to the outer disk and metal abundances are enhanced in that region, 
with the consequence that the gradients flatten out. A rough
estimate for the O/H gradient variation in these models 
leads to a steepening rate of
$\chi \simeq 0.0040-0.0060\, {\rm dex}\, {\rm kpc}^{-1}\, {\rm Gyr}^{-1}$,
which is consistent with our present results. A similar
behaviour has also been obtained by Alib\'es et al. 
(\cite{alibes}). On the other hand, models such as 
those based on two infall episodes by Chiappini et al. 
(\cite{cmr2001}), lead to some steepening of the gradients, 
even though the inside-out approach is adopted. Possibly,
the main reason for the different predictions
of the quoted models appears to reside on the different adopted
timescales for star formation and infall rate, so that we
expect our present results may be helpful in order to
constrain these quantities.

\begin{acknowledgements}
      This work was partially supported by FAPESP and CNPq.
\end{acknowledgements}

\end{document}